\documentclass[aps,prc,twocolumn,groupedaddress]{revtex4-1}

\usepackage{hyperref}
\usepackage{graphicx}
\usepackage{amsmath}
\usepackage{amssymb}
\graphicspath{ {images/} }

\newcommand{\sNN}{\sqrt{s_\mathrm{NN}}}
\newcommand{\JmidHat}{{\hat{J}_\mathrm{mid}}}
\newcommand{\JpartHat}{{\hat{J}_\mathrm{part}}}
\newcommand{\JspecHat}{{\hat{J}_\mathrm{spec}}}
\newcommand{\JsystHat}{{\hat{J}_\mathrm{syst}}}
\newcommand{\Jmid}{{\vec{J}_\mathrm{mid}}}
\newcommand{\Jpart}{{\vec{J}_\mathrm{part}}}
\newcommand{\Jspec}{{\vec{J}_\mathrm{spec}}}
\newcommand{\Jsyst}{{\vec{J}_\mathrm{syst}}}
\newcommand{\SystPartCorrelator}{\JpartHat\cdot\JsystHat}
\newcommand{\SpecPartCorrelator}{\JpartHat\cdot\JspecHat}

\newcommand{\SpecMidCorrelator}{\JmidHat\cdot\JspecHat}
\newcommand{\CorrelatorPhiTwoPart}{\sin(\Phi_2-\phi_\JpartHat)}
\newcommand{\CorrelatorPhiTwoMid}{\sin(\Phi_2-\phi_\JmidHat)}
\newcommand{\HadronSpin}{\vec{S}_{\rm H}}
\newcommand{\PHyper}{\overline{P}_\mathrm{H}}

\begin{document}

\title{Decorrelation of participant and spectator angular momenta in heavy-ion collisions}

\author{Joseph R. Adams, Michael A. Lisa}
\affiliation{Department of Physics, Ohio State University, Columbus, Ohio 43210 USA}

\date{\today}

\begin{abstract}
High-energy heavy-ion collisions contain enormous angular momentum, $|\vec{J}|$, which is $\mathcal{O}(10^3-10^6\hbar)$ in the range of
 collision energy, $\sNN$, spanned experimentally by the Relativistic
 Heavy Ion Collider (RHIC) and the Large Hadron
 Collider (LHC). A fraction of $\vec{J}$ is transferred
 to the overlapping collision region, which is indispensable
 for measuring observables such as vorticity-driven hadron spin
 alignment with $\hat{J}$. Experiments estimate the orientation of
 $\hat{J}$ of the participant nucleons within the collision
 overlap region, $\JpartHat$, by using that of the
 forward- and backward-going spectating nucleons $\JspecHat$. Using two
 models, we study the decorrelation between $\JpartHat$ and
 $\JspecHat$, driven both by angular-momentum conservation and event-by-event
 fluctuations, as well as by the decorrelation between
 the orientation of the elliptic overlap region and
 the $\JpartHat$. $\sNN$-dependent decorrelation is observed in both
 of these cases and is large enough to
 be an important corrective factor used when experimentally observing phenomena driven by $\vec{J}$.
\end{abstract}

\maketitle

\section{Introduction}
Relativistic heavy-ion collisions generate sufficient energy densities to
 momentarily [$\mathcal{O}$(1 fm/$c$)] deconfine constituent quarks, and thereby
 create the conditions required to study the strong
 nuclear force described by quantum chromodynamics (QCD). The
 state of matter produced is the so-called quark-gluon
 plasma (QGP)~\cite{Shuryak:1980tp,Adams:2005dq,Adcox:2004mh,Back:2004je,Arsene:2004fa}, which has been the subject of
 intense study for decades; theorists and experimentalists study
 a variety of phenomena over a wide range
 of center-of-mass, nucleon-nucleon collision energy, $\sNN$, in an
 effort to characterize the QCD phase diagram~\cite{Akiba:2015jwa}. 

Phenomena driven by the angular momentum of participating nucleons within the collision overlap region, $\Jpart$,
 have been of much interest~\cite{Becattini:2020ngo,Liang:2004ph,Becattini:2008fmr,Betz:2007kg,Abelev:2007zk,STAR:2017ckg,Acharya:2019ryw,Adam:2020pti,Jiang:2016woz}. Experimentally, the orientation
 of $\JpartHat$ is not known exactly and must
 be approximated with the orientation of the angular
 momentum of the forward- and backward-going ``spectator" nucleons,
 $\JspecHat$, which are characterized by having a ``forward"
 rapidity, $y$, or pseudo-rapidity, $\eta$, with $|y|, |\eta|\gtrsim1-2$.
 The QGP state, formed in the collision overlap region,
 ultimately emits particles across a range
 of rapidity, but typically only the mid-rapidity ($|y|,
 |\eta|\lesssim1-2$) particles' paths are able to be reconstructed
 experimentally. Observables such as the spin orientations of
 hadrons, $\HadronSpin$, which have been shown to be
 globally aligned with $\JsystHat$~\cite{STAR:2017ckg,Acharya:2019ryw,Adam:2020pti,STAR:2021beb}, are therefore only accessible
 at mid-rapidity. Ideally, then, the correlation between observables
 such as $\HadronSpin$ and the angular-momentum orientation of
 the corresponding mid-rapidity region, $\JmidHat$, could be studied.

The angular momentum of the system, $\Jsyst$, is perpendicular in the transverse plane (which is orthogonal
 to the beam axis) to the the impact
 parameter, $\vec{b}$, connecting the centers of the two
 nuclei. $\JpartHat$ instead fluctuates about $\JsystHat$ on an
 event-by-event basis, due to the randomness of nucleon
 positions within the colliding nuclei. While the effects
 of initial-state fluctuations in heavy-ion collisions have been
 thoroughly investigated\cite{Alver:2010gr,Bzdak:2011yy,Bloczynski:2012en,Vovchenko:2013viu,Qin:2010pf,Lacey:2010hw,Qian:2016pau,Jia:2014jca,Luzum:2013yya,Magdy:2020gxf,Gardim:2020sma,Bhalerao:2019uzw,Ma:2016hkg}, it has yet to be discussed
 in the context of decorrelation between $\JpartHat$ and
 $\JspecHat$. 

Through the use of a simple and intuitive model, we
 show in this study that the effects of
 both initial-state fluctuations and angular-momentum conservation lead to
 a significant decorrelation between the true $\JpartHat$ and
 $\JpartHat$ as estimated by $\JspecHat$. Furthermore, by using
 a more realistic model which evolves in time
 and simulates parton interactions, we find a significantly
 larger suppression of the correlation between $\JspecHat$ and
 the $\JmidHat$ that is experimentally of interest. This
 decorrelation is strongly dependent on $\sNN$ and would
 require experimental observations of phenomena driven by angular
 momentum to correct for this effect. While it
 may seem natural to attempt to avoid such
 corrective factors by measuring $\JmidHat$ directly from the
 azimuthal distribution of particles emitted at mid-rapidity, we
 find a decorrelation between the orientation of the
 roughly elliptic shape of the overlap region, practically
 disallowing such a method.

\section{Models}
The first of two models used to study these effects is a simple Monte-Carlo Glauber (MCG) model~\cite{Miller:2007ri}. Our MCG model consists of randomly generating angular and radial coordinates according to the Woods-Saxon distribution,
\begin{equation}
\ \rho(r) = \frac{\rho_{0}}{1 + e^{\frac{r - R}{a}}} ,
\end{equation}
 with the appropriate Jacobians. In this study, we look at ${}^{197}_{97}$Au, where $\rho_0 = 0.1693$ fm${}^{-3}$, $R = 6.38$ fm, and $a = 0.535$ fm~\cite{LandoltBornstein2004:sm_lbs_978-3-540-45555-4_81}. An impact parameter $|\vec{b}|$ is chosen according to $dN/db\propto b$, and two nuclei are generated around two points a distance $b_{i}$ from each other before re-centering the nuclei to maintain the chosen $b$. While generating nucleon positions, any newly generated nucleon whose center lies within 0.9 fm of another nucleon's center within the same nucleus is regenerated. 

\begin{figure}[!h]
\centering
\includegraphics[width=\linewidth]{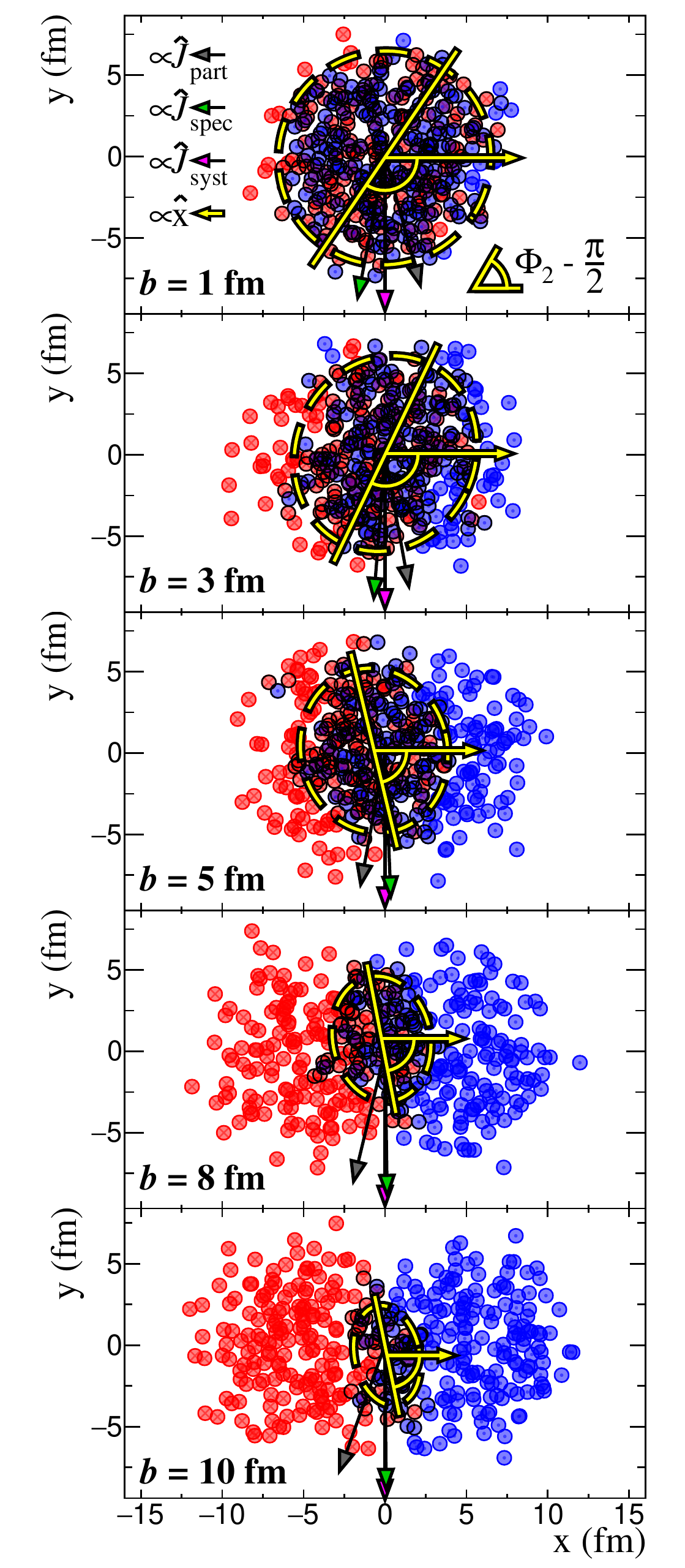}
\caption{\label{fig:coll} A series of collisions generated using our
 MCG model at varying b, viewed in the
 transverse plane. The participants are outlined in black
 and the elliptic fit to their positions is
 displayed in dashed yellow with a line through
 the major axis. The events shown are typical,
 in that $\SpecPartCorrelator|_b\approx\left<\SpecPartCorrelator\right>\Big\rvert_b$ and
  $|\sin(\Phi_2)|\Big\rvert_b\approx\left<|\sin(\Phi_2)|\right>\Big\rvert_b$.
  }
\end{figure}
\begin{figure}[!h]
\centering
\includegraphics[width=\linewidth]{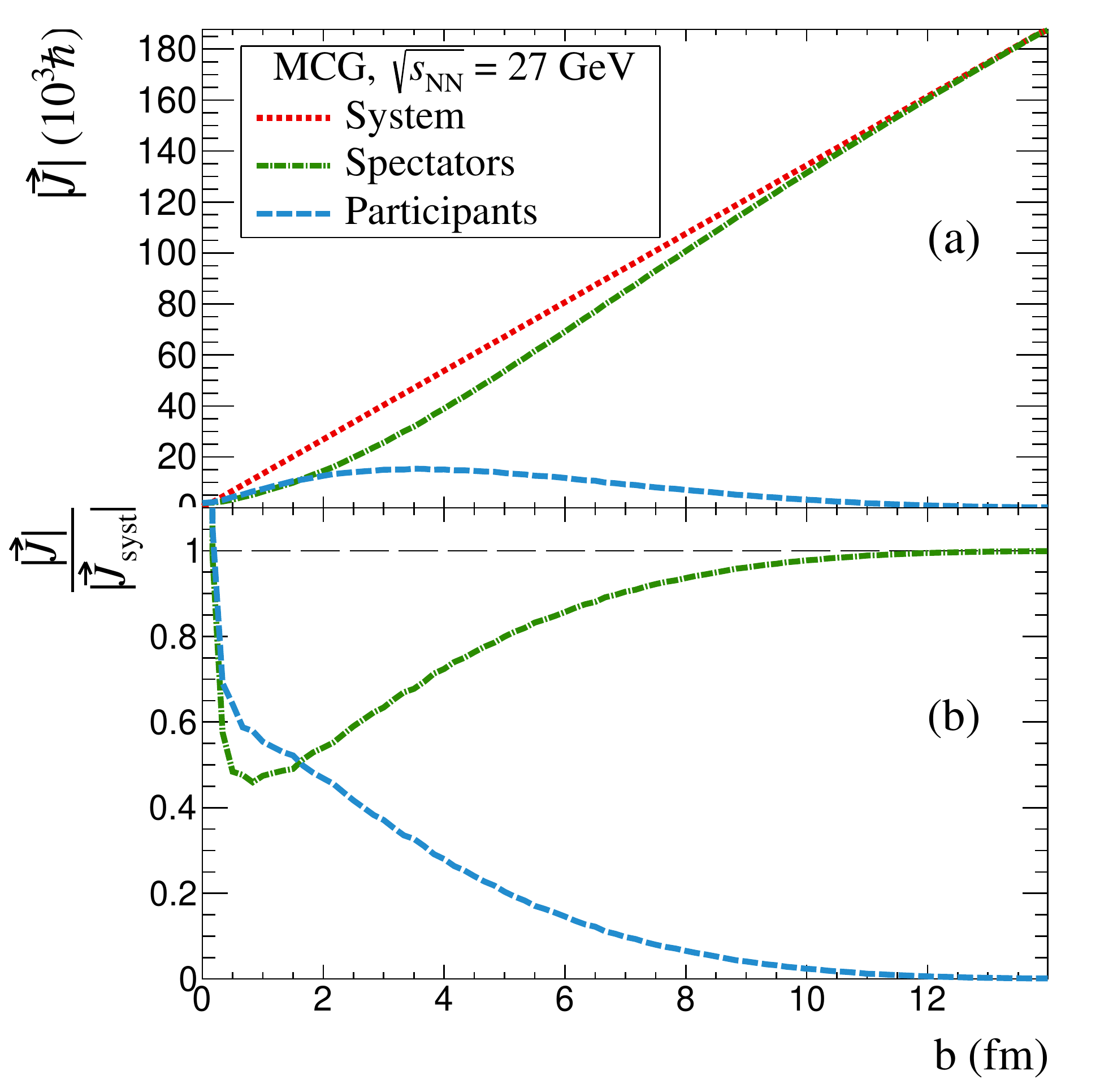}
\caption{\label{fig:b_AvgLMag_Mcg} $|\Jpart|$ is the largest at $b\approx3.5$ fm,
 but is sizable for all but peripheral collisions.
 For very central collisions ($b\lesssim0.5$ fm), $|\Jpart|$ and
 $|\Jspec|$ each become larger than $|\Jsyst|$ since spectators
 still exist while $|\Jsyst|\rightarrow0$ as $b\rightarrow0$.}
\end{figure}

The qualification that a nucleon must satisfy to be considered a participant is that its
 center must lie within $d_{\perp}$ of at least
 one nucleon from the other nucleus. $d_{\perp}$ is
 related to the beam energy and the nucleon-nucleon
 inelastic cross section at that energy, which is
 parameterized according to~\cite{Montanet:1994xu}:
\begin{multline}
\ \sigma^\mathrm{total}_\mathrm{NN} = 48 + 0.522 (\ln{p})^{2} - 4.51 \ln{p}
\\ \sigma_\mathrm{NN}^\mathrm{elastic} = 11.9 + 26.9 p^{-1.21} + 0.169 (\ln{p})^{2} - 1.85 \ln{p}
\\ d_{\perp} = \sqrt{\frac{\sigma_\mathrm{NN}^\mathrm{total} - \sigma_\mathrm{NN}^\mathrm{elastic}}{\pi}} = \sqrt{\frac{\sigma_\mathrm{NN}^\mathrm{inelastic}}{\pi}},
\end{multline}
where $p$ is the center-of-mass nucleon momentum.
The dependence of $d_\perp$ on $\sNN$ is weak.
For the collision energy $\sqrt{s_\mathrm{{NN}}} = 27$ GeV, somewhat arbitrarily chosen for our calculations, $d_{\perp} = 0.984$ fm.

A series of typical peripheral collisions using the
 MCG model at varying $b$ is shown in
 Fig.~\ref{fig:coll}, where $\Phi_2$ describes the 
 orientation of the collision and is defined in 
 Eq.~\ref{eq:Phi2}. The directions
 of $\JsystHat$, shown as magenta
 (dark gray) arrows, $\JpartHat$, shown as the gray 
 (medium gray) arrows, and $\JspecHat$, shown 
 as the green (light gray) arrows, are shown for
 each collision. $\JsystHat$
 always points in the $-\hat{y}$ direction while 
 $\JpartHat$ and $\JspecHat$ fluctuate about the $-\hat{y}$ 
 direction and point on opposite sides of $\JsystHat$. 
 While our MCG model calculations serve as
 a nice baseline for building intuitions, they do
 not incorporate any time evolution of the system
 and therefore do not allow us to study
 the effects of angular-momentum redistribution through particle interactions
 or to select regions in rapidity. 

As angular-momentum-driven phenomena are interested mainly in the QGP phase,
 we do not want to concern ourselves with
 late-stage interactions or decays which will act with
 a disproportionately large lever arm on the angular
 momentum of the region. The string-melting version of
 the a multi-phase transport" (AMPT) model~\cite{Lin:2004en} provides the
 ideal environment for a study of these angular
 momentum correlations with a more realistic description, while
 still allowing the user to ignore late-stage interactions
 and decays. The user has access to the
 positions and momenta of the spectators and of
 the partons at hadronization. Although hadronization will redistribute
 angular momentum to some degree, this is a
 sub-dominant effect.

AMPT uses the heavy-ion jet interaction generator (HIJING)~\cite{Gyulassy:1994ew} for initial
 conditions and Zhang's parton cascade (ZPC)~\cite{Zhang:1997ej} for handling
 partonic interactions. The Lund string fragmentation model is
 used for hadronization and a relativistic transport (ART)
 model is used for treating hadronic scatterings. For
 angular momentum calculations, we are only interested in
 the state of partons at the moment of
 hadronization, which is at the end of the
 ZPC stage.

The input parameters to AMPT, besides $\sNN$, the
 range of $b$, and the number of collisions,
 are not changed. For MCG and AMPT, 50K
 events are generated with $0\leq b\leq14$ fm. Only
 one collision energy is studied in MCG as
 the only energy-dependent effect is a slight reduction
 in $r_\perp$ with $\sNN$; the somewhat arbitrary choice
 is $\sNN=27$ GeV.

\section{$\Jpart, \Jmid$ Correlations with $\JsystHat, \JspecHat$\label{sec:CorrelationsWithJ}}

\begin{figure}
    \centering
    \includegraphics[width=\linewidth]{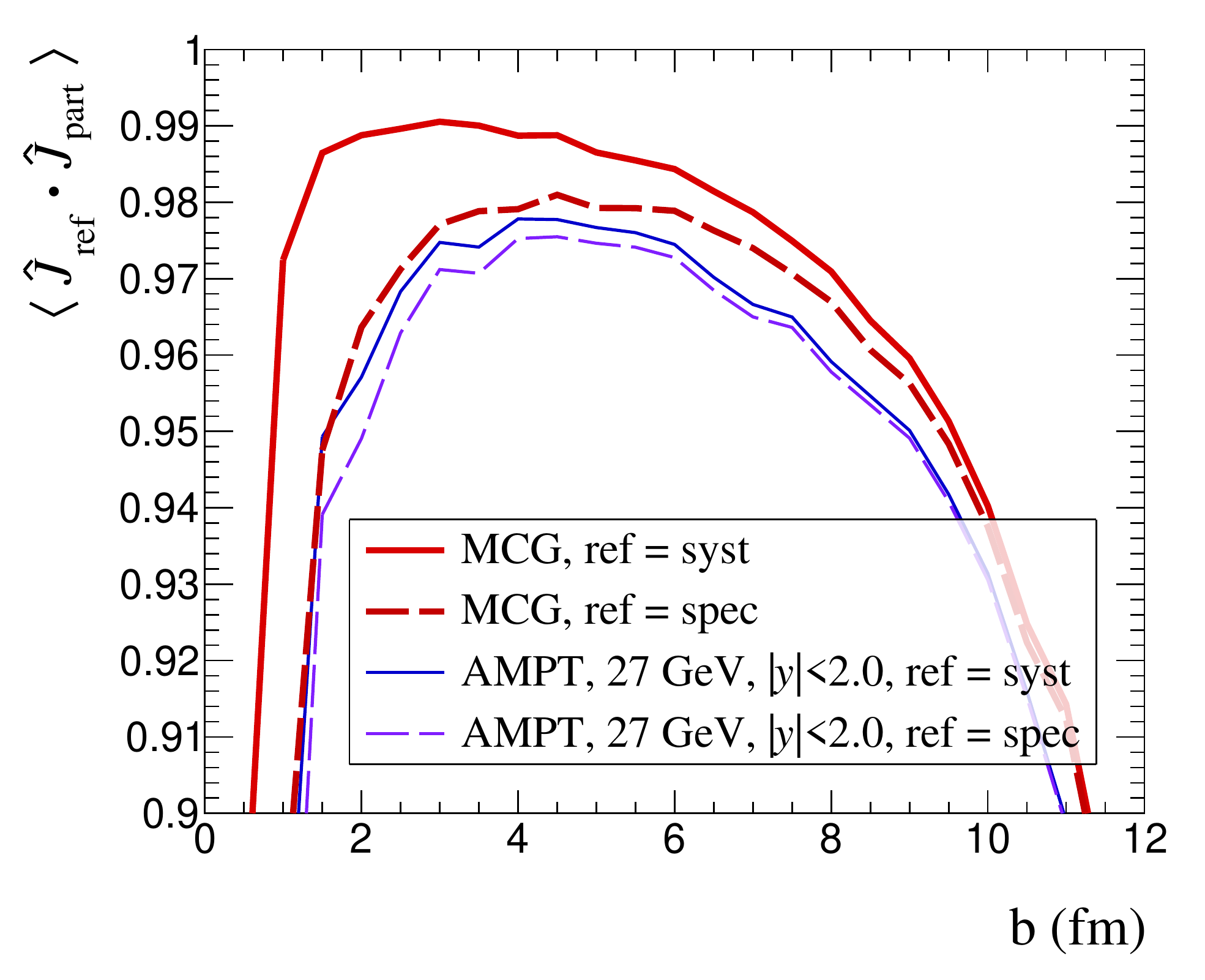}
    \caption{$\JpartHat$ and $\JsystHat$ are poorly correlated for central
 and peripheral collisions but are well correlated in
 between. The correlation between $\JpartHat$ and $\JspecHat$ is
 smaller than the correlation between $\JpartHat$ and $\JsystHat$,
 an effect driven by conservation of angular momentum.
 In AMPT, shown as the lower two lines, we choose 
 spectators with the somewhat
 arbitrary cut $|y|>2$ and in this figure the
 remaining particles are considered participants.}
    \label{fig:b_AvgLRefHatDotLPartHat}
\end{figure}

In central heavy-ion collisions ($b\lesssim 3$ fm), $|\Jsyst|$ and $|\Jpart|$ become smaller
 as $b\rightarrow0$ while $|\Jspec|$ remains large; many spectators
 still exist in these collisions (as seen for
 example in Fig.~\ref{fig:coll}), which carry a large lever
 arm. Because of this, $|\Jspec|$ and $|\Jpart|$ are
 non-zero even as $b\rightarrow0$ (see Fig.~\ref{fig:b_AvgLMag_Mcg}). As $b\rightarrow0$,
 $\JspecHat$ becomes more random and so, therefore, does
 $\JpartHat$ due to conservation of angular momentum. In
 peripheral collisions ($b\gtrsim9$ fm) $|\Jspec|$ dominates the contribution
 to $|\Jsyst|$ but the effects of initial-state fluctuations
 on $\JspecHat$ diminish as the number of spectators
 increases, so $\JspecHat$ becomes well aligned with $\Jsyst$;
 however, the number of participants drops as does
 the contribution of $|\Jpart|$ to $|\Jsyst|$, so initial-state
 fluctuations play a significant role in the orientation
 of $\JpartHat$. We might therefore expect $\JpartHat$ and
 $\JsystHat$ to be poorly correlated in central and
 peripheral collisions. In mid-central collisions, however, there are
 enough of both participants and spectators that initial-state
 fluctuations play a small role in the orientations
 of $\JpartHat$ and $\JspecHat$ and we might therefore expect them to be well correlated in these collisions.
We indeed see this behavior in the solid
 lines in Fig.~\ref{fig:b_AvgLRefHatDotLPartHat} measuring $\SystPartCorrelator$ with the MCG
 and AMPT models. Here and henceforth we use
 the rapidity cut $|y|>2$ in AMPT to approximately
 isolate the spectators, as would be done experimentally.
 When choosing the upper limit of $|y|=2$ to
 define the participant region in AMPT, no particles
 are excluded and we therefore see quite good
 agreement between the two models.

When instead measuring the correlation between the participants and
 the spectators, we see that $\SpecPartCorrelator<\SystPartCorrelator$;
 this is true both on average as well
 as event by event, and must be so
 because of conservation of angular momentum. This is
 represented in Fig.~\ref{fig:coll} as a cartoon of mid-central
 collisions within the MCG model viewed in the
 transverse plane. By design, $\JsystHat||-\hat{y}$ but initial-state fluctuations
 will generate a deviation of $\JpartHat$($\JspecHat$) from $-\hat{y}$
 and because of angular-momentum conservation $\JspecHat$($\JpartHat$) must point
 along the ``other side" of $-\hat{y}$; i.e., the
 angle between $\JpartHat$ and $\JspecHat$ must be larger
 than the angle between $\JpartHat$ and $\JsystHat$. 

\begin{figure}
    \centering
    \includegraphics[width=\linewidth]{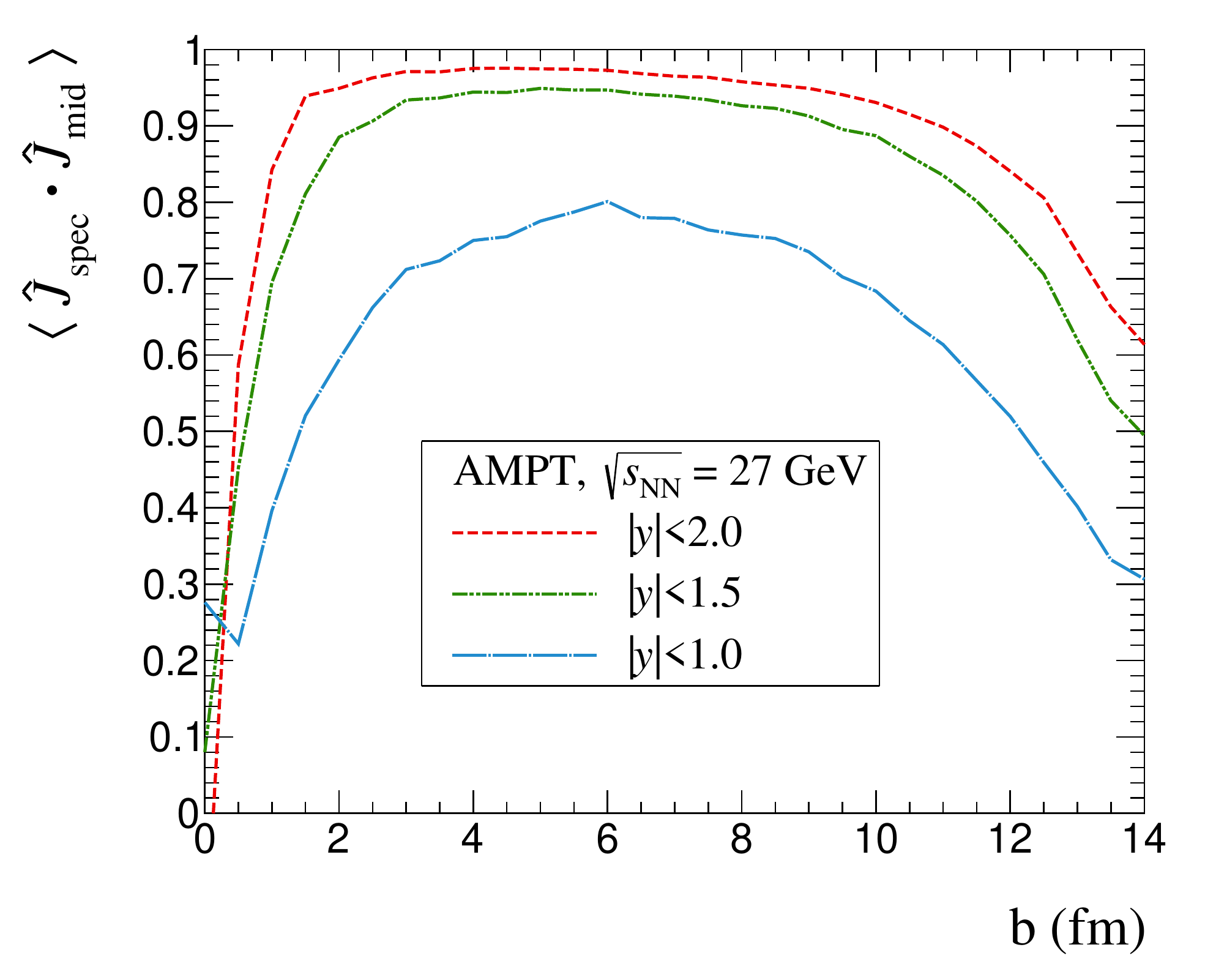}
    \caption{The correlation between $\JmidHat$ and
 $\JspecHat$ becomes smaller as we further constrain the
 size of the rapidity window used for the
 calculation of $\JmidHat$ where initial-state fluctuations play a
 larger role. Experiments typically are limited to $|y|<1$.
 Recall that $|y|=2$ is our participant-spectator cutoff in
 AMPT, so that $\vec{J}_{\mathrm{mid}, |y|<2}=\Jpart$.}
    \label{fig:b_AvgLSpecHatDotLPartHat_Rapidities}
\end{figure}

Experiments are typically set up to identify particles with tracking at mid-rapidity (e.g., with time
 projection chambers) while particle-type-insensitive detectors are placed at
 forward and backward rapidities to measure particle ``hits"
 (e.g., with calorimeters). When measuring phenomena driven by
 angular momentum within the QGP (e.g. global $\HadronSpin$
 alignment with $\hat{J}$), QGP byproducts are reconstructed at
 mid-rapidity while $\Jspec$ is measured using the azimuthal
 distribution of forward-/backward-going particles as an approximation of
 $\Jpart$; however, such an approximation is subject not
 only to the effects seen in Fig.~\ref{fig:b_AvgLRefHatDotLPartHat} but
 also to the experimental constraints of incomplete detector
 coverage and imperfect detector efficiencies. Because of this
 limitation, random fluctuations will play a larger role
 and we might expect the correlation between $\JspecHat$
 and $\JmidHat$ to be smaller than the correlation
 between $\JspecHat$ and $\JpartHat$.
This effect is shown in Fig.~\ref{fig:b_AvgLSpecHatDotLPartHat_Rapidities} within the
 AMPT model. For mid-central collisions, 
 $\JmidHat$ is well aligned with $\JsystHat$ when
 considering $|y|<2$; however, the degree of alignment drops
 substantially when considering the region $|y|<1$ typically used
 in experimental studies. This is striking; if taken
 at face value, this would translate to a
 correction of roughly 25\% on the observable of
 interest.

\begin{figure}
    \centering
    \includegraphics[width=\linewidth]{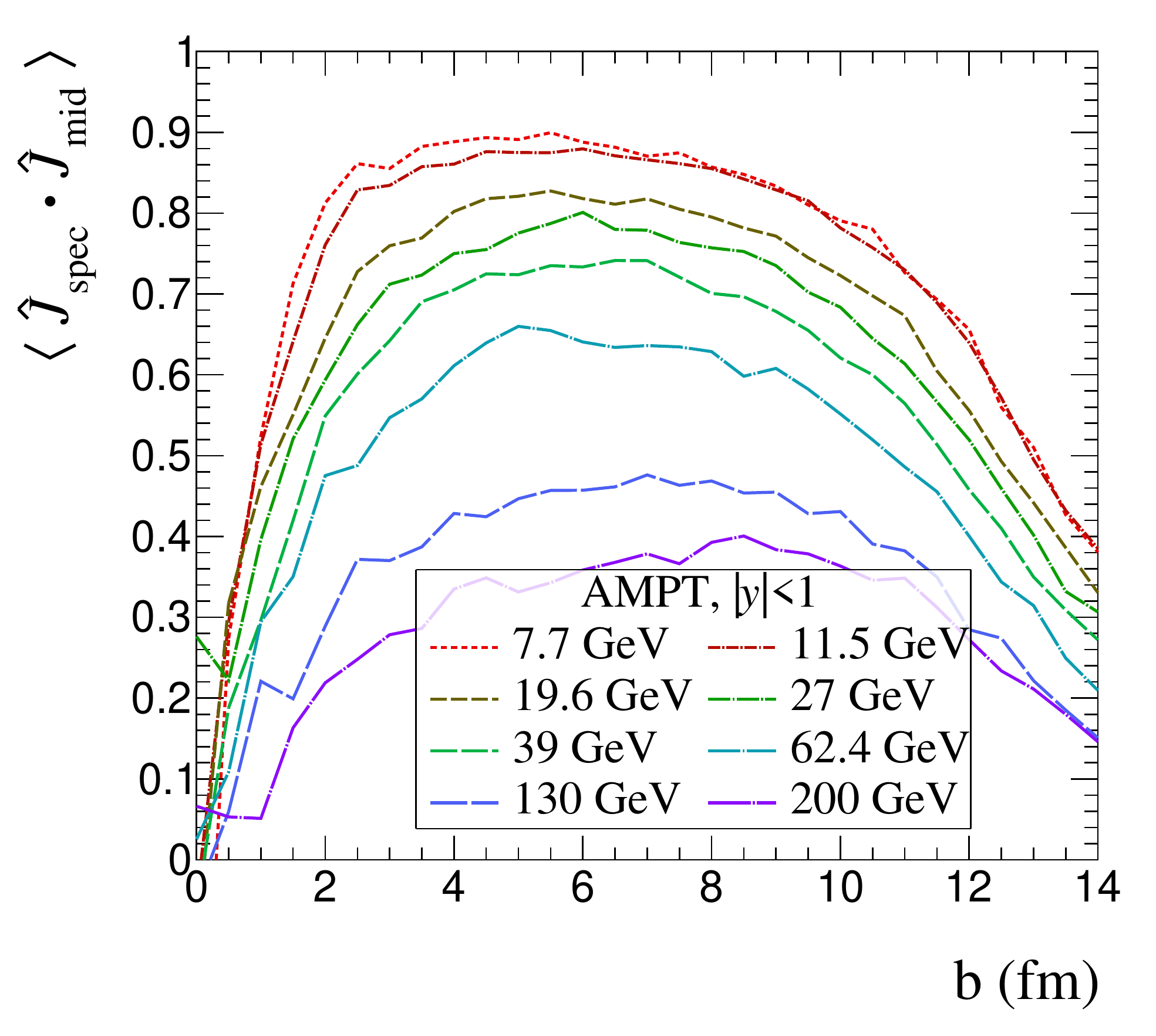}
    \caption{The correlation between $\JmidHat$ and
 $\JspecHat$ becomes smaller as we increase $\sNN$, where
 a given rapidity window includes a smaller fraction
 of emitted particles and initial-state fluctuations play a
 larger role. This is similar to the effects
 driving the observation in Fig.~\ref{fig:b_AvgLSpecHatDotLPartHat_Rapidities}. The values of
 $\sNN$ are chosen to match those of the
 RHIC Beam Energy Scan (BES).}
    \label{fig:b_AvgLSpecHatDotLPartHat_Energies}
\end{figure}

\begin{figure}
    \centering
    \includegraphics[width=\linewidth]{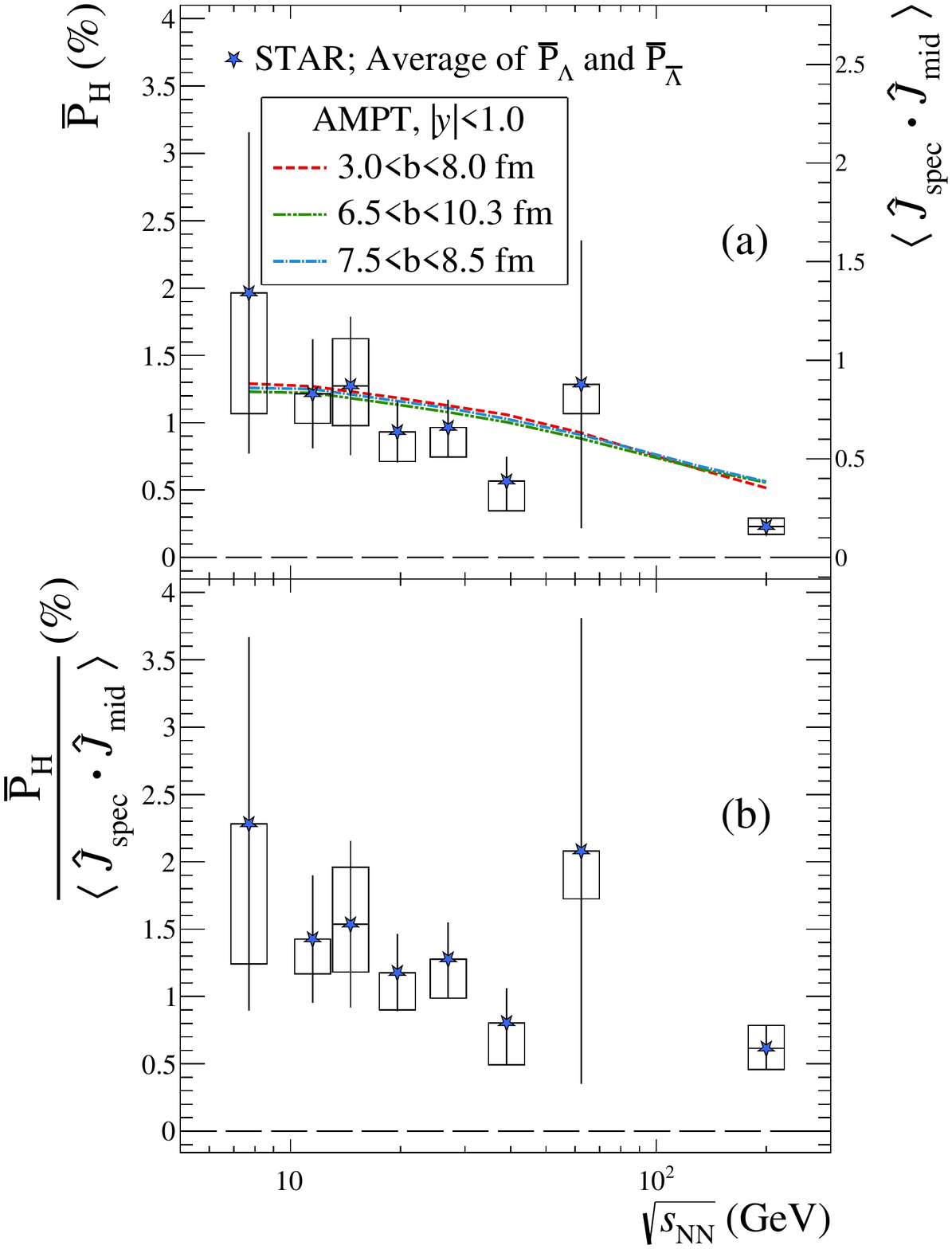}
    \caption{Panel (a) shows that the correlation between $\JmidHat$ and $\JspecHat$ for mid-central collisions
 (the event class used when studying angular-momentum-driven phenomena)
 falls with $\sNN$. $3<b<8$ fm describes the region
 from Fig.~\ref{fig:b_AvgLSpecHatDotLPartHat_Energies} where the correlation is flat, and
 $6.5<b<10.3$ fm and $7.5<b<8.5$ fm are two ways
 of approximating 20-50\% central collisions. Experimental results of
 global hyperon polarization~\cite{Abelev:2007zk,STAR:2017ckg,Adam:2018ivw}, $\PHyper$, are shown alongside these
 calculations. Also shown, in panel (b), are
 these results scaled by the relevant correlation calculated
 with AMPT.}
    \label{fig:sNN_PolarAndCorrelationAndTheirRatio}
\end{figure}

At larger collision energies, the fraction of emitted
 particles that lie in the rapidity window $|y|<1$
 becomes smaller; we might therefore expect the correlator
 $\left<\SpecMidCorrelator\right>$ to become smaller with increasing $\sNN$. On
 the other hand, the fraction of emitted particles
 that lie in the spectator (forward) rapidity region
 becomes larger and will impact $\JspecHat$. In Fig.~\ref{fig:b_AvgLSpecHatDotLPartHat_Energies}
 we see that, despite this, $\left<\SpecMidCorrelator\right>$ becomes smaller
 as $\sNN$ becomes larger. This correlation depends strongly
 on $\sNN$, differing by more than a factor
 of 2 between the lowest and highest collision
 energies. In Fig.~\ref{fig:sNN_PolarAndCorrelationAndTheirRatio}, $\left<\SpecMidCorrelator\right>$ is shown for mid-central
 collisions, defined in a number of ways that
 yield very similar results, as a function of
 $\sNN$. Any $\sNN$-dependent experimental observable driven by angular
 momentum within the QGP would be corrected in
 such a manner, by $\left<\SpecMidCorrelator\right>^{-1}$. Similarly, it is
 important for model predictions to use $\JpartHat$ or
 $\JmidHat$, rather than $\JsystHat$, when calculating phenomena driven
 by angular momentum within the QGP. 

 The global spin polarization of hyperons, $\PHyper$, is one
 such observable that would require correction, and experimental
 measurements of $\PHyper$~\cite{Abelev:2007zk,STAR:2017ckg,Adam:2018ivw} are
 shown alongside the correlator $\left<\SpecMidCorrelator\right>$ in Fig.~\ref{fig:sNN_PolarAndCorrelationAndTheirRatio}. Both
 $\PHyper$ and $\left<\SpecMidCorrelator\right>$ fall with increasing $\sNN$, though
 $\PHyper$ demonstrates a stronger dependence. The ratio of
 $\PHyper$ to $\left<\SpecMidCorrelator\right>$ is shown in the bottom
 panel of Fig.~\ref{fig:sNN_PolarAndCorrelationAndTheirRatio}; the scaled $\PHyper$ demonstrates a
 notably weaker dependence on $\sNN$.  In practice,
 a more detailed study, specific to a detector's
 coverage and acceptance or to a model's assumptions,
 would need to be performed to apply a
 correction. Without performing such corrections, there will be an
 apparent dependence on $\sNN$ driven at least in
 part by the behavior observed in Fig.~\ref{fig:sNN_PolarAndCorrelationAndTheirRatio}.

\section{$\JpartHat, \JmidHat$Correlations with geometry}
The overlap region of a heavy-ion collision is roughly elliptic on average, with the major axis of the
 ellipse somewhat aligned with $\hat{y}$.  We can fit the participant coordinates to an
 ellipse in order to determine its orientation and
 study the correlation between $\JpartHat$ and this orientation.
 We characterize the initial shape through harmonic-eccentricity
 coefficients $\varepsilon_{n}$ and event-plane angles $\Phi_{n}$ \cite{Qiu:2012uy}:
\begin{equation}
\varepsilon_{n} {e}^{in\Phi_{n}} = -\frac{ \int rdrd\phi  {r}^n {e}^{in\phi} e(r,\phi)}{\int rdrd\phi {r}^n e(r,\phi)} .
\end{equation}
By taking $n=2$ and treating the initial energy-density distribution $e(r,\phi)$
 as a sum of $\delta$ functions, each at the position of a nucleon, this reduces to
\begin{equation}\label{eq:Phi2}
\Phi_{2} = \frac{1}{2}\left[\arctan \frac{\sum_{i} {r_{\perp, i}}^2 \sin( 2 \phi_{i} )}{\sum_{i}{r_{\perp, i}}^2 \cos( 2 \phi_{i} ) }+\pi\right],
\end{equation}
where $r_{\perp}$ and $\phi$ are the polar coordinates
 of the participant nucleons in the transverse plane,
 as measured from the center of mass of
 the participants. This procedure is only applicable in
 the MCG model where all nucleons are either
 considered to be participants or spectators. In the
 AMPT model, we can reconstruct the orientation of
 the elliptic overlap region responsible for the mid-rapidity
 region by taking advantage of ``elliptic flow"; the
 pressure gradient is larger along the shorter axis
 of the ellipse than it is along the
 longer axis. Because of this, the azimuthal distribution
 of emitted particles in a rapidity window will
 reveal the orientation of the relevant overlap region~\cite{Voloshin:2008dg}:
\begin{equation}
    \Phi_2 = \frac{1}{2}\mathrm{atan2}\left(\sum_iw_i\sin(2\phi_i),\sum_iw_i\cos(2\phi_i)\right),
\end{equation}
where the weight, $w_i$, is typically the transverse momentum, $p_\mathrm{T}$. 

\begin{figure}
    \centering
    \includegraphics[width=\linewidth]{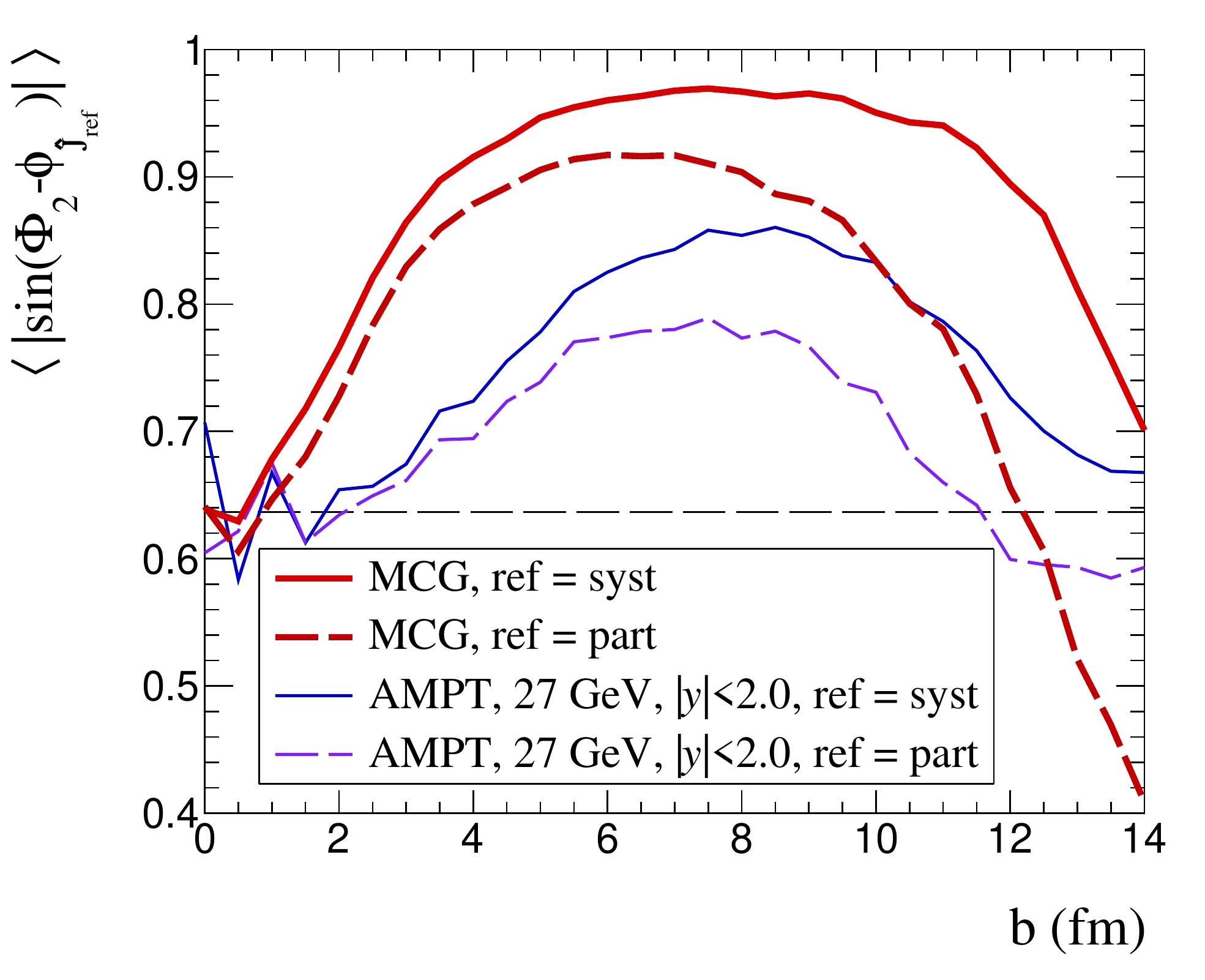}
    \caption{The correlation between the orientation 
    of the elliptic overlap region and $\JsystHat$ is 
    largest for mid-central collisions, in line with 
    expectations. Counter-intuitively, however, 
    there is a suppressed correlation between the 
    orientation of the ellipse and $\JpartHat$. 
    The absolute value of $\CorrelatorPhiTwoPart$ 
    is used since $\Phi_2$ is physically indistinguishable 
    from $\Phi_2\pm\pi$. Recall that $|y|=2$ is our 
    participant-spectator cutoff in AMPT, shown as the 
    lower two lines, so that 
    $\vec{J}_{\mathrm{mid}, |y|<2}=\Jpart$. The 
    horizontal dashed line represents $2/\pi$, which is 
    the average absolute value of the sine of the 
    difference between two random, uncorrelated numbers.}
    \label{fig:b_AvgSinPhi2MinusPhiLRef}
\end{figure}

For central collisions, the overlap region is quite
 circular and for very peripheral collisions only a
 small number of nucleons participate; in both cases,
 initial-state fluctuations play a large role in the
 orientation of the elliptic fit and therefore on
 $\Phi_2$. In mid-central collisions, the overlap region is
 sufficiently elliptic and there are enough participants that
 initial-state fluctuations will be sub-dominant; we might therefore
 expect $\Phi_2$ to be best aligned with $\phi_\JsystHat\pm\pi/2$
 for mid-central collisions. Such behavior is apparent in
 the solid lines in Fig.~\ref{fig:b_AvgSinPhi2MinusPhiLRef}.

We might also intuitively make the na\"{i}ve assumption that the somewhat elliptic
 participant region would be spinning about its major
 axis and therefore expect better alignment between $\Phi_2$
 and $\phi_\JpartHat, \phi_\JmidHat$ than between $\Phi_2$ and $\phi_\JsystHat$.
 If this were true, then the problematic suppression
 of $\SpecMidCorrelator$ discussed in Sec.~\ref{sec:CorrelationsWithJ} could potentially be
 avoided by measuring $\JmidHat$ directly from $\Phi_2$; however,
 when considering again Fig.~\ref{fig:b_AvgSinPhi2MinusPhiLRef}, there is apparently a
 significant suppression of the correlation between  $\Phi_2$
 and $\phi_\JpartHat, \phi_\JmidHat$. This can be understood by
 dividing a given tilted elliptic overlap region in
 two, lengthwise, and considering that one half is
 dominated by positive-rapidity nucleons while the other is dominated
 by negative-rapidity nucleons. By applying the right-hand rule
 to these two halves it is clear that
 $\JpartHat$ will tilt to the left as the
 elliptic overlap region tilts to the right, and
 vice versa. One can see in Fig.~\ref{fig:coll} a
 few examples of the major axis tilting away
 from $\JpartHat$ over a range of $b$.

For the same reasons that we expected $\SpecMidCorrelator$
 to decrease both when reducing the size of
 the window in $y$ and when increasing $\sNN$,
 we might expect the correlation $\left<\left|\CorrelatorPhiTwoMid\right|\right>$ to again
 decrease in AMPT when considering $|y|<1$, as well
 as when considering larger $\sNN$. While not shown
 in this paper, we indeed did find such
 additional suppressions to this correlation. This decorrelation between
 $\Phi_2$ and $\phi_\JpartHat, \phi_\JmidHat$ demonstrates that one can
 not avoid the corrective factors discussed in Sec.~\ref{sec:CorrelationsWithJ}
 by indirectly measuring $\JmidHat$ through $\Phi_2$.

\section{Summary}
Initial-state fluctuations drive a decorrelation between $\JsystHat$ and
 $\JpartHat, \JmidHat$, which is the largest for central
 and peripheral collisions. Conservation of angular momentum further
 suppresses this correlation between $\JspecHat$ and $\JpartHat, \JmidHat$,
 albeit slightly. Only $\JspecHat$ is experimentally accessible, and
 is used as an approximation of $\JmidHat$. As
 the size of the mid-rapidity window becomes smaller,
 the correlation between $\JspecHat$ and $\JmidHat$ is suppressed
 further. Similarly, this decorrelation becomes more dramatic with
 increasing $\sNN$. 

The orientation of the elliptic overlap region, $\Phi_2$, has a smaller
 correlation with $\JpartHat, \JmidHat$ than with $\JsystHat$, in
 conflict with potentially intuitive expectations. The correlation between
 the orientation of elliptic shape and $\JpartHat$ is
 further suppressed when constraining the mid-rapidity window to
 $|y|<1$, as well as when increasing $\sNN$. Deducing
 $\JpartHat$ from $\Phi_2$ is therefore not a viable
 method to avoid the problems of correlation suppression
 between $\JmidHat$ and $\JspecHat$.

The findings presented here hold significant implications for measurements of phenomena
 driven by angular momentum within the QGP, and
 particularly for those interested in the dependence on
 $\sNN$. Based on our model-dependent study, it is
 crucial for studies of angular-momentum-driven phenomena, such as
 experimental measurements of $\PHyper$, to correct
 for the decorrelation between $\JspecHat$ and $\JmidHat$ in
 a $\sNN$-dependent manner, and for model predictions of
 such phenomena to use $\JmidHat$ instead of $\JsystHat$.
 Without these corrections, any observed dependence will be driven at
 least in part by this decorrelation.

\section{Acknowledgements}
We thank Jinfeng Liao and Giorgio Torrieri for helpful conversations. 
This work was supported by the U.S. Department of Energy grant DE-SC0020651.

\newpage
\bibliography{AngularMomentumDecorrelation}

\providecommand{\noopsort}[1]{}\providecommand{\singleletter}[1]{#1}%
\begin{thebibliography}{38}%
\makeatletter
\providecommand \@ifxundefined [1]{%
 \@ifx{#1\undefined}
}%
\providecommand \@ifnum [1]{%
 \ifnum #1\expandafter \@firstoftwo
 \else \expandafter \@secondoftwo
 \fi
}%
\providecommand \@ifx [1]{%
 \ifx #1\expandafter \@firstoftwo
 \else \expandafter \@secondoftwo
 \fi
}%
\providecommand \natexlab [1]{#1}%
\providecommand \enquote  [1]{``#1''}%
\providecommand \bibnamefont  [1]{#1}%
\providecommand \bibfnamefont [1]{#1}%
\providecommand \citenamefont [1]{#1}%
\providecommand \href@noop [0]{\@secondoftwo}%
\providecommand \href [0]{\begingroup \@sanitize@url \@href}%
\providecommand \@href[1]{\@@startlink{#1}\@@href}%
\providecommand \@@href[1]{\endgroup#1\@@endlink}%
\providecommand \@sanitize@url [0]{\catcode `\\12\catcode `\$12\catcode
  `\&12\catcode `\#12\catcode `\^12\catcode `\_12\catcode `\%12\relax}%
\providecommand \@@startlink[1]{}%
\providecommand \@@endlink[0]{}%
\providecommand \url  [0]{\begingroup\@sanitize@url \@url }%
\providecommand \@url [1]{\endgroup\@href {#1}{\urlprefix }}%
\providecommand \urlprefix  [0]{URL }%
\providecommand \Eprint [0]{\href }%
\providecommand \doibase [0]{http://dx.doi.org/}%
\providecommand \selectlanguage [0]{\@gobble}%
\providecommand \bibinfo  [0]{\@secondoftwo}%
\providecommand \bibfield  [0]{\@secondoftwo}%
\providecommand \translation [1]{[#1]}%
\providecommand \BibitemOpen [0]{}%
\providecommand \bibitemStop [0]{}%
\providecommand \bibitemNoStop [0]{.\EOS\space}%
\providecommand \EOS [0]{\spacefactor3000\relax}%
\providecommand \BibitemShut  [1]{\csname bibitem#1\endcsname}%
\let\auto@bib@innerbib\@empty
\bibitem [{\citenamefont {Shuryak}(1980)}]{Shuryak:1980tp}%
  \BibitemOpen
  \bibfield  {author} {\bibinfo {author} {\bibfnamefont {E.~V.}\ \bibnamefont
  {Shuryak}},\ }\href {\doibase 10.1016/0370-1573(80)90105-2} {\bibfield
  {journal} {\bibinfo  {journal} {Phys. Rept.}\ }\textbf {\bibinfo {volume}
  {61}},\ \bibinfo {pages} {71} (\bibinfo {year} {1980})}\BibitemShut {NoStop}%
\bibitem [{\citenamefont {Adams}\ \emph {et~al.}(2005)\citenamefont {Adams}
  \emph {et~al.}}]{Adams:2005dq}%
  \BibitemOpen
  \bibfield  {author} {\bibinfo {author} {\bibfnamefont {J.}~\bibnamefont
  {Adams}} \emph {et~al.} (\bibinfo {collaboration} {STAR}),\ }\href {\doibase
  10.1016/j.nuclphysa.2005.03.085} {\bibfield  {journal} {\bibinfo  {journal}
  {Nucl. Phys. A}\ }\textbf {\bibinfo {volume} {757}},\ \bibinfo {pages} {102}
  (\bibinfo {year} {2005})},\ \Eprint {http://arxiv.org/abs/nucl-ex/0501009}
  {arXiv:nucl-ex/0501009} \BibitemShut {NoStop}%
\bibitem [{\citenamefont {Adcox}\ \emph {et~al.}(2005)\citenamefont {Adcox}
  \emph {et~al.}}]{Adcox:2004mh}%
  \BibitemOpen
  \bibfield  {author} {\bibinfo {author} {\bibfnamefont {K.}~\bibnamefont
  {Adcox}} \emph {et~al.} (\bibinfo {collaboration} {PHENIX}),\ }\href
  {\doibase 10.1016/j.nuclphysa.2005.03.086} {\bibfield  {journal} {\bibinfo
  {journal} {Nucl. Phys. A}\ }\textbf {\bibinfo {volume} {757}},\ \bibinfo
  {pages} {184} (\bibinfo {year} {2005})},\ \Eprint
  {http://arxiv.org/abs/nucl-ex/0410003} {arXiv:nucl-ex/0410003} \BibitemShut
  {NoStop}%
\bibitem [{\citenamefont {Back}\ \emph {et~al.}(2005)\citenamefont {Back} \emph
  {et~al.}}]{Back:2004je}%
  \BibitemOpen
  \bibfield  {author} {\bibinfo {author} {\bibfnamefont {B.~B.}\ \bibnamefont
  {Back}} \emph {et~al.} (\bibinfo {collaboration} {PHOBOS}),\ }\href {\doibase
  10.1016/j.nuclphysa.2005.03.084} {\bibfield  {journal} {\bibinfo  {journal}
  {Nucl. Phys. A}\ }\textbf {\bibinfo {volume} {757}},\ \bibinfo {pages} {28}
  (\bibinfo {year} {2005})},\ \Eprint {http://arxiv.org/abs/nucl-ex/0410022}
  {arXiv:nucl-ex/0410022} \BibitemShut {NoStop}%
\bibitem [{\citenamefont {Arsene}\ \emph {et~al.}(2005)\citenamefont {Arsene}
  \emph {et~al.}}]{Arsene:2004fa}%
  \BibitemOpen
  \bibfield  {author} {\bibinfo {author} {\bibfnamefont {I.}~\bibnamefont
  {Arsene}} \emph {et~al.} (\bibinfo {collaboration} {BRAHMS}),\ }\href
  {\doibase 10.1016/j.nuclphysa.2005.02.130} {\bibfield  {journal} {\bibinfo
  {journal} {Nucl. Phys. A}\ }\textbf {\bibinfo {volume} {757}},\ \bibinfo
  {pages} {1} (\bibinfo {year} {2005})},\ \Eprint
  {http://arxiv.org/abs/nucl-ex/0410020} {arXiv:nucl-ex/0410020} \BibitemShut
  {NoStop}%
\bibitem [{\citenamefont {Akiba}\ \emph {et~al.}(2015)\citenamefont {Akiba}
  \emph {et~al.}}]{Akiba:2015jwa}%
  \BibitemOpen
  \bibfield  {author} {\bibinfo {author} {\bibfnamefont {Y.}~\bibnamefont
  {Akiba}} \emph {et~al.},\ }\href@noop {} {\  (\bibinfo {year} {2015})},\
  \Eprint {http://arxiv.org/abs/1502.02730} {arXiv:1502.02730 [nucl-ex]}
  \BibitemShut {NoStop}%
\bibitem [{\citenamefont {Becattini}\ and\ \citenamefont
  {Lisa}(2020)}]{Becattini:2020ngo}%
  \BibitemOpen
  \bibfield  {author} {\bibinfo {author} {\bibfnamefont {F.}~\bibnamefont
  {Becattini}}\ and\ \bibinfo {author} {\bibfnamefont {M.~A.}\ \bibnamefont
  {Lisa}},\ }\href {\doibase 10.1146/annurev-nucl-021920-095245} {\bibfield
  {journal} {\bibinfo  {journal} {Ann. Rev. Nucl. Part. Sci.}\ }\textbf
  {\bibinfo {volume} {70}},\ \bibinfo {pages} {395} (\bibinfo {year} {2020})},\
  \Eprint {http://arxiv.org/abs/2003.03640} {arXiv:2003.03640 [nucl-ex]}
  \BibitemShut {NoStop}%
\bibitem [{\citenamefont {Liang}\ and\ \citenamefont
  {Wang}(2005)}]{Liang:2004ph}%
  \BibitemOpen
  \bibfield  {author} {\bibinfo {author} {\bibfnamefont {Z.-T.}\ \bibnamefont
  {Liang}}\ and\ \bibinfo {author} {\bibfnamefont {X.-N.}\ \bibnamefont
  {Wang}},\ }\href {\doibase 10.1103/PhysRevLett.94.102301} {\bibfield
  {journal} {\bibinfo  {journal} {Phys. Rev. Lett.}\ }\textbf {\bibinfo
  {volume} {94}},\ \bibinfo {pages} {102301} (\bibinfo {year} {2005})},\
  \bibinfo {note} {[Erratum: Phys.Rev.Lett. 96, 039901 (2006)]},\ \Eprint
  {http://arxiv.org/abs/nucl-th/0410079} {arXiv:nucl-th/0410079} \BibitemShut
  {NoStop}%
\bibitem [{\citenamefont {Becattini}\ and\ \citenamefont
  {Piccinini}(2008)}]{Becattini:2008fmr}%
  \BibitemOpen
  \bibfield  {author} {\bibinfo {author} {\bibfnamefont {F.}~\bibnamefont
  {Becattini}}\ and\ \bibinfo {author} {\bibfnamefont {F.}~\bibnamefont
  {Piccinini}},\ }\href@noop {} {\bibfield  {journal} {\bibinfo  {journal} {J.
  Phys. G}\ }\textbf {\bibinfo {volume} {35}},\ \bibinfo {pages} {054001.155}
  (\bibinfo {year} {2008})}\BibitemShut {NoStop}%
\bibitem [{\citenamefont {Betz}\ \emph {et~al.}(2007)\citenamefont {Betz},
  \citenamefont {Gyulassy},\ and\ \citenamefont {Torrieri}}]{Betz:2007kg}%
  \BibitemOpen
  \bibfield  {author} {\bibinfo {author} {\bibfnamefont {B.}~\bibnamefont
  {Betz}}, \bibinfo {author} {\bibfnamefont {M.}~\bibnamefont {Gyulassy}}, \
  and\ \bibinfo {author} {\bibfnamefont {G.}~\bibnamefont {Torrieri}},\ }\href
  {\doibase 10.1103/PhysRevC.76.044901} {\bibfield  {journal} {\bibinfo
  {journal} {Phys. Rev. C}\ }\textbf {\bibinfo {volume} {76}},\ \bibinfo
  {pages} {044901} (\bibinfo {year} {2007})},\ \Eprint
  {http://arxiv.org/abs/0708.0035} {arXiv:0708.0035 [nucl-th]} \BibitemShut
  {NoStop}%
\bibitem [{\citenamefont {Abelev}\ \emph {et~al.}(2007)\citenamefont {Abelev}
  \emph {et~al.}}]{Abelev:2007zk}%
  \BibitemOpen
  \bibfield  {author} {\bibinfo {author} {\bibfnamefont {B.~I.}\ \bibnamefont
  {Abelev}} \emph {et~al.} (\bibinfo {collaboration} {STAR}),\ }\href {\doibase
  10.1103/PhysRevC.76.024915} {\bibfield  {journal} {\bibinfo  {journal} {Phys.
  Rev. C}\ }\textbf {\bibinfo {volume} {76}},\ \bibinfo {pages} {024915}
  (\bibinfo {year} {2007})},\ \bibinfo {note} {[Erratum: Phys.Rev.C 95, 039906
  (2017)]},\ \Eprint {http://arxiv.org/abs/0705.1691} {arXiv:0705.1691
  [nucl-ex]} \BibitemShut {NoStop}%
\bibitem [{\citenamefont {Adamczyk}\ \emph {et~al.}(2017)\citenamefont
  {Adamczyk} \emph {et~al.}}]{STAR:2017ckg}%
  \BibitemOpen
  \bibfield  {author} {\bibinfo {author} {\bibfnamefont {L.}~\bibnamefont
  {Adamczyk}} \emph {et~al.} (\bibinfo {collaboration} {STAR}),\ }\href
  {\doibase 10.1038/nature23004} {\bibfield  {journal} {\bibinfo  {journal}
  {Nature}\ }\textbf {\bibinfo {volume} {548}},\ \bibinfo {pages} {62}
  (\bibinfo {year} {2017})},\ \Eprint {http://arxiv.org/abs/1701.06657}
  {arXiv:1701.06657 [nucl-ex]} \BibitemShut {NoStop}%
\bibitem [{\citenamefont {Acharya}\ \emph {et~al.}(2020)\citenamefont {Acharya}
  \emph {et~al.}}]{Acharya:2019ryw}%
  \BibitemOpen
  \bibfield  {author} {\bibinfo {author} {\bibfnamefont {S.}~\bibnamefont
  {Acharya}} \emph {et~al.} (\bibinfo {collaboration} {ALICE}),\ }\href
  {\doibase 10.1103/PhysRevC.101.044611} {\bibfield  {journal} {\bibinfo
  {journal} {Phys. Rev. C}\ }\textbf {\bibinfo {volume} {101}},\ \bibinfo
  {pages} {044611} (\bibinfo {year} {2020})},\ \Eprint
  {http://arxiv.org/abs/1909.01281} {arXiv:1909.01281 [nucl-ex]} \BibitemShut
  {NoStop}%
\bibitem [{\citenamefont {Adam}\ \emph {et~al.}(2021)\citenamefont {Adam} \emph
  {et~al.}}]{Adam:2020pti}%
  \BibitemOpen
  \bibfield  {author} {\bibinfo {author} {\bibfnamefont {J.}~\bibnamefont
  {Adam}} \emph {et~al.} (\bibinfo {collaboration} {STAR}),\ }\href {\doibase
  10.1103/PhysRevLett.126.162301} {\bibfield  {journal} {\bibinfo  {journal}
  {Phys. Rev. Lett.}\ }\textbf {\bibinfo {volume} {126}},\ \bibinfo {pages}
  {162301} (\bibinfo {year} {2021})},\ \Eprint
  {http://arxiv.org/abs/2012.13601} {arXiv:2012.13601 [nucl-ex]} \BibitemShut
  {NoStop}%
\bibitem [{\citenamefont {Jiang}\ \emph {et~al.}(2016)\citenamefont {Jiang},
  \citenamefont {Lin},\ and\ \citenamefont {Liao}}]{Jiang:2016woz}%
  \BibitemOpen
  \bibfield  {author} {\bibinfo {author} {\bibfnamefont {Y.}~\bibnamefont
  {Jiang}}, \bibinfo {author} {\bibfnamefont {Z.-W.}\ \bibnamefont {Lin}}, \
  and\ \bibinfo {author} {\bibfnamefont {J.}~\bibnamefont {Liao}},\ }\href
  {\doibase 10.1103/PhysRevC.94.044910} {\bibfield  {journal} {\bibinfo
  {journal} {Phys. Rev. C}\ }\textbf {\bibinfo {volume} {94}},\ \bibinfo
  {pages} {044910} (\bibinfo {year} {2016})},\ \bibinfo {note} {[Erratum:
  Phys.Rev.C 95, 049904 (2017)]},\ \Eprint {http://arxiv.org/abs/1602.06580}
  {arXiv:1602.06580 [hep-ph]} \BibitemShut {NoStop}%
\bibitem [{\citenamefont {Abdallah}\ \emph {et~al.}(2021)\citenamefont
  {Abdallah} \emph {et~al.}}]{STAR:2021beb}%
  \BibitemOpen
  \bibfield  {author} {\bibinfo {author} {\bibfnamefont {M.~S.}\ \bibnamefont
  {Abdallah}} \emph {et~al.} (\bibinfo {collaboration} {STAR}),\ }\href
  {\doibase 10.1103/PhysRevC.104.L061901} {\bibfield  {journal} {\bibinfo
  {journal} {Phys. Rev. C}\ }\textbf {\bibinfo {volume} {104}},\ \bibinfo
  {pages} {L061901} (\bibinfo {year} {2021})},\ \Eprint
  {http://arxiv.org/abs/2108.00044} {arXiv:2108.00044 [nucl-ex]} \BibitemShut
  {NoStop}%
\bibitem [{\citenamefont {Alver}\ and\ \citenamefont
  {Roland}(2010)}]{Alver:2010gr}%
  \BibitemOpen
  \bibfield  {author} {\bibinfo {author} {\bibfnamefont {B.}~\bibnamefont
  {Alver}}\ and\ \bibinfo {author} {\bibfnamefont {G.}~\bibnamefont {Roland}},\
  }\href {\doibase 10.1103/PhysRevC.82.039903} {\bibfield  {journal} {\bibinfo
  {journal} {Phys. Rev. C}\ }\textbf {\bibinfo {volume} {81}},\ \bibinfo
  {pages} {054905} (\bibinfo {year} {2010})},\ \bibinfo {note} {[Erratum:
  Phys.Rev.C 82, 039903 (2010)]},\ \Eprint {http://arxiv.org/abs/1003.0194}
  {arXiv:1003.0194 [nucl-th]} \BibitemShut {NoStop}%
\bibitem [{\citenamefont {Bzdak}\ and\ \citenamefont
  {Skokov}(2012)}]{Bzdak:2011yy}%
  \BibitemOpen
  \bibfield  {author} {\bibinfo {author} {\bibfnamefont {A.}~\bibnamefont
  {Bzdak}}\ and\ \bibinfo {author} {\bibfnamefont {V.}~\bibnamefont {Skokov}},\
  }\href {\doibase 10.1016/j.physletb.2012.02.065} {\bibfield  {journal}
  {\bibinfo  {journal} {Phys. Lett. B}\ }\textbf {\bibinfo {volume} {710}},\
  \bibinfo {pages} {171} (\bibinfo {year} {2012})},\ \Eprint
  {http://arxiv.org/abs/1111.1949} {arXiv:1111.1949 [hep-ph]} \BibitemShut
  {NoStop}%
\bibitem [{\citenamefont {Bloczynski}\ \emph {et~al.}(2013)\citenamefont
  {Bloczynski}, \citenamefont {Huang}, \citenamefont {Zhang},\ and\
  \citenamefont {Liao}}]{Bloczynski:2012en}%
  \BibitemOpen
  \bibfield  {author} {\bibinfo {author} {\bibfnamefont {J.}~\bibnamefont
  {Bloczynski}}, \bibinfo {author} {\bibfnamefont {X.-G.}\ \bibnamefont
  {Huang}}, \bibinfo {author} {\bibfnamefont {X.}~\bibnamefont {Zhang}}, \ and\
  \bibinfo {author} {\bibfnamefont {J.}~\bibnamefont {Liao}},\ }\href {\doibase
  10.1016/j.physletb.2012.12.030} {\bibfield  {journal} {\bibinfo  {journal}
  {Phys. Lett. B}\ }\textbf {\bibinfo {volume} {718}},\ \bibinfo {pages} {1529}
  (\bibinfo {year} {2013})},\ \Eprint {http://arxiv.org/abs/1209.6594}
  {arXiv:1209.6594 [nucl-th]} \BibitemShut {NoStop}%
\bibitem [{\citenamefont {Vovchenko}\ \emph {et~al.}(2013)\citenamefont
  {Vovchenko}, \citenamefont {Anchishkin},\ and\ \citenamefont
  {Csernai}}]{Vovchenko:2013viu}%
  \BibitemOpen
  \bibfield  {author} {\bibinfo {author} {\bibfnamefont {V.}~\bibnamefont
  {Vovchenko}}, \bibinfo {author} {\bibfnamefont {D.}~\bibnamefont
  {Anchishkin}}, \ and\ \bibinfo {author} {\bibfnamefont {L.~P.}\ \bibnamefont
  {Csernai}},\ }\href {\doibase 10.1103/PhysRevC.88.014901} {\bibfield
  {journal} {\bibinfo  {journal} {Phys. Rev. C}\ }\textbf {\bibinfo {volume}
  {88}},\ \bibinfo {pages} {014901} (\bibinfo {year} {2013})},\ \Eprint
  {http://arxiv.org/abs/1306.5208} {arXiv:1306.5208 [nucl-th]} \BibitemShut
  {NoStop}%
\bibitem [{\citenamefont {Qin}\ \emph {et~al.}(2010)\citenamefont {Qin},
  \citenamefont {Petersen}, \citenamefont {Bass},\ and\ \citenamefont
  {Muller}}]{Qin:2010pf}%
  \BibitemOpen
  \bibfield  {author} {\bibinfo {author} {\bibfnamefont {G.-Y.}\ \bibnamefont
  {Qin}}, \bibinfo {author} {\bibfnamefont {H.}~\bibnamefont {Petersen}},
  \bibinfo {author} {\bibfnamefont {S.~A.}\ \bibnamefont {Bass}}, \ and\
  \bibinfo {author} {\bibfnamefont {B.}~\bibnamefont {Muller}},\ }\href
  {\doibase 10.1103/PhysRevC.82.064903} {\bibfield  {journal} {\bibinfo
  {journal} {Phys. Rev. C}\ }\textbf {\bibinfo {volume} {82}},\ \bibinfo
  {pages} {064903} (\bibinfo {year} {2010})},\ \Eprint
  {http://arxiv.org/abs/1009.1847} {arXiv:1009.1847 [nucl-th]} \BibitemShut
  {NoStop}%
\bibitem [{\citenamefont {Lacey}\ \emph {et~al.}(2011)\citenamefont {Lacey},
  \citenamefont {Wei}, \citenamefont {Ajitanand},\ and\ \citenamefont
  {Taranenko}}]{Lacey:2010hw}%
  \BibitemOpen
  \bibfield  {author} {\bibinfo {author} {\bibfnamefont {R.~A.}\ \bibnamefont
  {Lacey}}, \bibinfo {author} {\bibfnamefont {R.}~\bibnamefont {Wei}}, \bibinfo
  {author} {\bibfnamefont {N.~N.}\ \bibnamefont {Ajitanand}}, \ and\ \bibinfo
  {author} {\bibfnamefont {A.}~\bibnamefont {Taranenko}},\ }\href {\doibase
  10.1103/PhysRevC.83.044902} {\bibfield  {journal} {\bibinfo  {journal} {Phys.
  Rev. C}\ }\textbf {\bibinfo {volume} {83}},\ \bibinfo {pages} {044902}
  (\bibinfo {year} {2011})},\ \Eprint {http://arxiv.org/abs/1009.5230}
  {arXiv:1009.5230 [nucl-ex]} \BibitemShut {NoStop}%
\bibitem [{\citenamefont {Qian}\ and\ \citenamefont
  {Heinz}(2016)}]{Qian:2016pau}%
  \BibitemOpen
  \bibfield  {author} {\bibinfo {author} {\bibfnamefont {J.}~\bibnamefont
  {Qian}}\ and\ \bibinfo {author} {\bibfnamefont {U.}~\bibnamefont {Heinz}},\
  }\href {\doibase 10.1103/PhysRevC.94.024910} {\bibfield  {journal} {\bibinfo
  {journal} {Phys. Rev. C}\ }\textbf {\bibinfo {volume} {94}},\ \bibinfo
  {pages} {024910} (\bibinfo {year} {2016})},\ \Eprint
  {http://arxiv.org/abs/1607.01732} {arXiv:1607.01732 [nucl-th]} \BibitemShut
  {NoStop}%
\bibitem [{\citenamefont {Jia}(2014)}]{Jia:2014jca}%
  \BibitemOpen
  \bibfield  {author} {\bibinfo {author} {\bibfnamefont {J.}~\bibnamefont
  {Jia}},\ }\href {\doibase 10.1088/0954-3899/41/12/124003} {\bibfield
  {journal} {\bibinfo  {journal} {J. Phys. G}\ }\textbf {\bibinfo {volume}
  {41}},\ \bibinfo {pages} {124003} (\bibinfo {year} {2014})},\ \Eprint
  {http://arxiv.org/abs/1407.6057} {arXiv:1407.6057 [nucl-ex]} \BibitemShut
  {NoStop}%
\bibitem [{\citenamefont {Luzum}\ and\ \citenamefont
  {Petersen}(2014)}]{Luzum:2013yya}%
  \BibitemOpen
  \bibfield  {author} {\bibinfo {author} {\bibfnamefont {M.}~\bibnamefont
  {Luzum}}\ and\ \bibinfo {author} {\bibfnamefont {H.}~\bibnamefont
  {Petersen}},\ }\href {\doibase 10.1088/0954-3899/41/6/063102} {\bibfield
  {journal} {\bibinfo  {journal} {J. Phys. G}\ }\textbf {\bibinfo {volume}
  {41}},\ \bibinfo {pages} {063102} (\bibinfo {year} {2014})},\ \Eprint
  {http://arxiv.org/abs/1312.5503} {arXiv:1312.5503 [nucl-th]} \BibitemShut
  {NoStop}%
\bibitem [{\citenamefont {Magdy}\ \emph {et~al.}(2020)\citenamefont {Magdy},
  \citenamefont {Sun}, \citenamefont {Ye}, \citenamefont {Evdokimov},\ and\
  \citenamefont {Lacey}}]{Magdy:2020gxf}%
  \BibitemOpen
  \bibfield  {author} {\bibinfo {author} {\bibfnamefont {N.}~\bibnamefont
  {Magdy}}, \bibinfo {author} {\bibfnamefont {X.}~\bibnamefont {Sun}}, \bibinfo
  {author} {\bibfnamefont {Z.}~\bibnamefont {Ye}}, \bibinfo {author}
  {\bibfnamefont {O.}~\bibnamefont {Evdokimov}}, \ and\ \bibinfo {author}
  {\bibfnamefont {R.}~\bibnamefont {Lacey}},\ }\href {\doibase
  10.3390/universe6090146} {\bibfield  {journal} {\bibinfo  {journal}
  {Universe}\ }\textbf {\bibinfo {volume} {6}},\ \bibinfo {pages} {146}
  (\bibinfo {year} {2020})},\ \Eprint {http://arxiv.org/abs/2009.02734}
  {arXiv:2009.02734 [nucl-ex]} \BibitemShut {NoStop}%
\bibitem [{\citenamefont {Gardim}\ \emph {et~al.}(2021)\citenamefont {Gardim},
  \citenamefont {Giacalone}, \citenamefont {Luzum},\ and\ \citenamefont
  {Ollitrault}}]{Gardim:2020sma}%
  \BibitemOpen
  \bibfield  {author} {\bibinfo {author} {\bibfnamefont {F.~G.}\ \bibnamefont
  {Gardim}}, \bibinfo {author} {\bibfnamefont {G.}~\bibnamefont {Giacalone}},
  \bibinfo {author} {\bibfnamefont {M.}~\bibnamefont {Luzum}}, \ and\ \bibinfo
  {author} {\bibfnamefont {J.-Y.}\ \bibnamefont {Ollitrault}},\ }\href
  {\doibase 10.1016/j.nuclphysa.2020.121999} {\bibfield  {journal} {\bibinfo
  {journal} {Nucl. Phys. A}\ }\textbf {\bibinfo {volume} {1005}},\ \bibinfo
  {pages} {121999} (\bibinfo {year} {2021})},\ \Eprint
  {http://arxiv.org/abs/2002.07008} {arXiv:2002.07008 [nucl-th]} \BibitemShut
  {NoStop}%
\bibitem [{\citenamefont {Bhalerao}\ \emph {et~al.}(2019)\citenamefont
  {Bhalerao}, \citenamefont {Giacalone}, \citenamefont
  {Guerrero-Rodr\'\i{}guez}, \citenamefont {Luzum}, \citenamefont {Marquet},\
  and\ \citenamefont {Ollitrault}}]{Bhalerao:2019uzw}%
  \BibitemOpen
  \bibfield  {author} {\bibinfo {author} {\bibfnamefont {R.~S.}\ \bibnamefont
  {Bhalerao}}, \bibinfo {author} {\bibfnamefont {G.}~\bibnamefont {Giacalone}},
  \bibinfo {author} {\bibfnamefont {P.}~\bibnamefont
  {Guerrero-Rodr\'\i{}guez}}, \bibinfo {author} {\bibfnamefont
  {M.}~\bibnamefont {Luzum}}, \bibinfo {author} {\bibfnamefont
  {C.}~\bibnamefont {Marquet}}, \ and\ \bibinfo {author} {\bibfnamefont
  {J.-Y.}\ \bibnamefont {Ollitrault}},\ }\href {\doibase
  10.5506/APhysPolB.50.1165} {\bibfield  {journal} {\bibinfo  {journal} {Acta
  Phys. Polon. B}\ }\textbf {\bibinfo {volume} {50}},\ \bibinfo {pages} {1165}
  (\bibinfo {year} {2019})},\ \Eprint {http://arxiv.org/abs/1903.06366}
  {arXiv:1903.06366 [nucl-th]} \BibitemShut {NoStop}%
\bibitem [{\citenamefont {Ma}\ \emph {et~al.}(2016)\citenamefont {Ma},
  \citenamefont {Ma},\ and\ \citenamefont {Ma}}]{Ma:2016hkg}%
  \BibitemOpen
  \bibfield  {author} {\bibinfo {author} {\bibfnamefont {L.}~\bibnamefont
  {Ma}}, \bibinfo {author} {\bibfnamefont {G.~L.}\ \bibnamefont {Ma}}, \ and\
  \bibinfo {author} {\bibfnamefont {Y.~G.}\ \bibnamefont {Ma}},\ }\href
  {\doibase 10.1103/PhysRevC.94.044915} {\bibfield  {journal} {\bibinfo
  {journal} {Phys. Rev. C}\ }\textbf {\bibinfo {volume} {94}},\ \bibinfo
  {pages} {044915} (\bibinfo {year} {2016})},\ \Eprint
  {http://arxiv.org/abs/1610.04733} {arXiv:1610.04733 [nucl-th]} \BibitemShut
  {NoStop}%
\bibitem [{\citenamefont {Miller}\ \emph {et~al.}(2007)\citenamefont {Miller},
  \citenamefont {Reygers}, \citenamefont {Sanders},\ and\ \citenamefont
  {Steinberg}}]{Miller:2007ri}%
  \BibitemOpen
  \bibfield  {author} {\bibinfo {author} {\bibfnamefont {M.~L.}\ \bibnamefont
  {Miller}}, \bibinfo {author} {\bibfnamefont {K.}~\bibnamefont {Reygers}},
  \bibinfo {author} {\bibfnamefont {S.~J.}\ \bibnamefont {Sanders}}, \ and\
  \bibinfo {author} {\bibfnamefont {P.}~\bibnamefont {Steinberg}},\ }\href
  {\doibase 10.1146/annurev.nucl.57.090506.123020} {\bibfield  {journal}
  {\bibinfo  {journal} {Ann. Rev. Nucl. Part. Sci.}\ }\textbf {\bibinfo
  {volume} {57}},\ \bibinfo {pages} {205} (\bibinfo {year} {2007})},\ \Eprint
  {http://arxiv.org/abs/nucl-ex/0701025} {arXiv:nucl-ex/0701025} \BibitemShut
  {NoStop}%
\bibitem [{\citenamefont {Fricke}\ and\ \citenamefont
  {Heilig}()}]{LandoltBornstein2004:sm_lbs_978-3-540-45555-4_81}%
  \BibitemOpen
  \bibfield  {author} {\bibinfo {author} {\bibfnamefont {G.}~\bibnamefont
  {Fricke}}\ and\ \bibinfo {author} {\bibfnamefont {K.}~\bibnamefont
  {Heilig}},\ }\href {\doibase 10.1007/10856314_81} {\enquote {\bibinfo {title}
  {Nuclear charge radii {\textperiodcentered} 79-au gold: Datasheet from
  landolt-b{\"o}rnstein - group i elementary particles, nuclei and atoms
  {\textperiodcentered} volume 20: ``nuclear charge radii'' in
  springermaterials (https://doi.org/10.1007/10856314{\_}81)},}\ }\bibinfo
  {note} {Copyright 2004 Springer-Verlag Berlin Heidelberg}\BibitemShut
  {NoStop}%
\bibitem [{\citenamefont {Montanet}\ \emph {et~al.}(1994)\citenamefont
  {Montanet} \emph {et~al.}}]{Montanet:1994xu}%
  \BibitemOpen
  \bibfield  {author} {\bibinfo {author} {\bibfnamefont {L.}~\bibnamefont
  {Montanet}} \emph {et~al.} (\bibinfo {collaboration} {Particle Data Group}),\
  }\href {\doibase 10.1103/PhysRevD.50.1173} {\bibfield  {journal} {\bibinfo
  {journal} {Phys. Rev. D}\ }\textbf {\bibinfo {volume} {50}},\ \bibinfo
  {pages} {1173} (\bibinfo {year} {1994})}\BibitemShut {NoStop}%
\bibitem [{\citenamefont {Lin}\ \emph {et~al.}(2005)\citenamefont {Lin},
  \citenamefont {Ko}, \citenamefont {Li}, \citenamefont {Zhang},\ and\
  \citenamefont {Pal}}]{Lin:2004en}%
  \BibitemOpen
  \bibfield  {author} {\bibinfo {author} {\bibfnamefont {Z.-W.}\ \bibnamefont
  {Lin}}, \bibinfo {author} {\bibfnamefont {C.~M.}\ \bibnamefont {Ko}},
  \bibinfo {author} {\bibfnamefont {B.-A.}\ \bibnamefont {Li}}, \bibinfo
  {author} {\bibfnamefont {B.}~\bibnamefont {Zhang}}, \ and\ \bibinfo {author}
  {\bibfnamefont {S.}~\bibnamefont {Pal}},\ }\href {\doibase
  10.1103/PhysRevC.72.064901} {\bibfield  {journal} {\bibinfo  {journal} {Phys.
  Rev. C}\ }\textbf {\bibinfo {volume} {72}},\ \bibinfo {pages} {064901}
  (\bibinfo {year} {2005})},\ \Eprint {http://arxiv.org/abs/nucl-th/0411110}
  {arXiv:nucl-th/0411110} \BibitemShut {NoStop}%
\bibitem [{\citenamefont {Gyulassy}\ and\ \citenamefont
  {Wang}(1994)}]{Gyulassy:1994ew}%
  \BibitemOpen
  \bibfield  {author} {\bibinfo {author} {\bibfnamefont {M.}~\bibnamefont
  {Gyulassy}}\ and\ \bibinfo {author} {\bibfnamefont {X.-N.}\ \bibnamefont
  {Wang}},\ }\href {\doibase 10.1016/0010-4655(94)90057-4} {\bibfield
  {journal} {\bibinfo  {journal} {Comput. Phys. Commun.}\ }\textbf {\bibinfo
  {volume} {83}},\ \bibinfo {pages} {307} (\bibinfo {year} {1994})},\ \Eprint
  {http://arxiv.org/abs/nucl-th/9502021} {arXiv:nucl-th/9502021} \BibitemShut
  {NoStop}%
\bibitem [{\citenamefont {Zhang}(1998)}]{Zhang:1997ej}%
  \BibitemOpen
  \bibfield  {author} {\bibinfo {author} {\bibfnamefont {B.}~\bibnamefont
  {Zhang}},\ }\href {\doibase 10.1016/S0010-4655(98)00010-1} {\bibfield
  {journal} {\bibinfo  {journal} {Comput. Phys. Commun.}\ }\textbf {\bibinfo
  {volume} {109}},\ \bibinfo {pages} {193} (\bibinfo {year} {1998})},\ \Eprint
  {http://arxiv.org/abs/nucl-th/9709009} {arXiv:nucl-th/9709009} \BibitemShut
  {NoStop}%
\bibitem [{\citenamefont {Adam}\ \emph {et~al.}(2018)\citenamefont {Adam} \emph
  {et~al.}}]{Adam:2018ivw}%
  \BibitemOpen
  \bibfield  {author} {\bibinfo {author} {\bibfnamefont {J.}~\bibnamefont
  {Adam}} \emph {et~al.} (\bibinfo {collaboration} {STAR}),\ }\href {\doibase
  10.1103/PhysRevC.98.014910} {\bibfield  {journal} {\bibinfo  {journal} {Phys.
  Rev. C}\ }\textbf {\bibinfo {volume} {98}},\ \bibinfo {pages} {014910}
  (\bibinfo {year} {2018})},\ \Eprint {http://arxiv.org/abs/1805.04400}
  {arXiv:1805.04400 [nucl-ex]} \BibitemShut {NoStop}%
\bibitem [{\citenamefont {Qiu}\ and\ \citenamefont {Heinz}(2012)}]{Qiu:2012uy}%
  \BibitemOpen
  \bibfield  {author} {\bibinfo {author} {\bibfnamefont {Z.}~\bibnamefont
  {Qiu}}\ and\ \bibinfo {author} {\bibfnamefont {U.}~\bibnamefont {Heinz}},\
  }\href {\doibase 10.1016/j.physletb.2012.09.030} {\bibfield  {journal}
  {\bibinfo  {journal} {Phys. Lett. B}\ }\textbf {\bibinfo {volume} {717}},\
  \bibinfo {pages} {261} (\bibinfo {year} {2012})},\ \Eprint
  {http://arxiv.org/abs/1208.1200} {arXiv:1208.1200 [nucl-th]} \BibitemShut
  {NoStop}%
\bibitem [{\citenamefont {Voloshin}\ \emph {et~al.}(2010)\citenamefont
  {Voloshin}, \citenamefont {Poskanzer},\ and\ \citenamefont
  {Snellings}}]{Voloshin:2008dg}%
  \BibitemOpen
  \bibfield  {author} {\bibinfo {author} {\bibfnamefont {S.~A.}\ \bibnamefont
  {Voloshin}}, \bibinfo {author} {\bibfnamefont {A.~M.}\ \bibnamefont
  {Poskanzer}}, \ and\ \bibinfo {author} {\bibfnamefont {R.}~\bibnamefont
  {Snellings}},\ }\href {\doibase 10.1007/978-3-642-01539-7_10} {\bibfield
  {journal} {\bibinfo  {journal} {Landolt-Bornstein}\ }\textbf {\bibinfo
  {volume} {23}},\ \bibinfo {pages} {293} (\bibinfo {year} {2010})},\ \Eprint
  {http://arxiv.org/abs/0809.2949} {arXiv:0809.2949 [nucl-ex]} \BibitemShut
  {NoStop}%
\end{thebibliography}%

\end{document}